\documentclass{pasj01}
\usepackage{mathrsfs} 
\usepackage{graphicx} 
\usepackage{color}  
 
\def\kms{km s$^{-1}$}
\def\vlsr{V_{\rm lsr}}

\def\ve{V_{\rm expa}}

\def\Msun{M_\odot} 
\def\deg{^\circ} 
\def\Tb{T_{\rm B}}  
\def\zhalf{h} 
\def\Hcc{ {\rm H\ cm^{-3}}}
\def\conex{\Delta_T}
\def\Tav{\langle T \rangle} 
\def\MG{ McClure-Griffiths } 
   
\title{Giant HI Hole inside the 3-kpc Ring and the North Polar Spur -- The Galactic Crater -- }  
\author{Yoshiaki \textsc{Sofue}\altaffilmark{1}  }
\altaffiltext{1}{Insitute of Astronomy, The University of Tokyo, Mitaka, Tokyo 181-0015, Japan } 
\email{sofue@ioa.s.u-tokyo.ac.jp}


\begin{document} 
\date{ } 
\maketitle  

\begin{abstract} 

Applying a newly developed tangent-circle method (TCM), we derive a volume density map of HI gas in the inner Galaxy as a function of galacto-centric distance $R$ and height $Z$. The HI hole around the Galactic Center (GC) is shown to have a crater-shaped wall, which coincides with the brightest ridge of the North Polar Spur and emanates from the 3-kpc expanding ring. The crater structure is explained by sweeping of the halo gas by a shock-wave from the GC. The unperturbed HI halo outside 3 kpc is shown to be in hydrostatic equilibrium, obeying the sech$^2 Z/h$ density law with a scale height $h \sim 450$ pc.
\end{abstract}

\section{Introduction} 
 
Among the number of expanding features in the Galactic disk, the most massive object is known as the 3-kpc ring in the HI- and CO-line observations (Cohen and Davies 1976; Bania 1977; Oort 1977 for review; Dame and Thaddeus 2008; Garcia et al. 2014). The near-side arm of the ring is approaching, and therefore, expanding at velocity of $\ve \sim 53$ \kms (Cohen and Davies 1976; Oort 1977), and the far-side is receding at $\ve \sim 56$ \kms (Dame and Thaddeus 2008). For such a motion the ring has been interpreted as an expanding shock front driven by an explosion at the Galactic Center (GC) (Sanders and Prendergast 1974; Oort 1977; Sofue 1976, 1977; 1984).  
Coinciding with the ring, a giant HI hole has been found in the halo around the  GC, which was interpreted as a void swept by the nuclear wind associated with the Fermi Bubbles (Lockman 1984; Lockman and \MG 2016).

As the Milky Way is evidenced to be a barred galaxy, the ring's motion has been more widely believed to be due to an oval flow in the bar potential (Binney et al. 1991; Rodriguez-Fernandez and Combes 2008). On the other hand, the explosion hypothesis has recently been highlighted according to accumulating evidences and arguments for energetic phenomena in the GC such as the high-temperature plasma (Uchiyama et al. 2013), bipolar hyper shells (BHS) in radio and X-rays (Sofue 1994, 2000, 2017;  Bland-Hawthorn and Cohen 2003; Sofue et al. 2016; and the literature therein for observations), and the Fermi Bubbles in $\gamma$-rays (Su et al. 2010; Bordoloi et al. 2017). Also, if the Galaxy is barred, the activity in the GC is inevitable as the consequence of bar-shocked gas accretion.

In this paper, we investigate the relationship of the HI hole discovered by Lockman and \MG (2016) with the 3-kpc HI ring and the North Polar Spur, and discuss their origin based on the GC explosion model.

\section{Radio and HI Spurs}   
 
Lockman and McClure-Griffiths (2016) showed that the HI hole has a clear wall at $R\sim 3$ kpc, recognized as a remarkable change of vertical extent in the channel maps at $\vlsr \sim \pm 130$ \kms at tangent longitude of $l\sim 20\deg$. 
The integrated intensity (column density) map along the tangent circle showed a bipolar void surrounding the Fermi Bubbles, and they explained the structure as due to galactic wind-driven HI hole.  
     
\begin{figure} 
\begin{center}    
\vskip -2mm
\includegraphics[width=8cm]{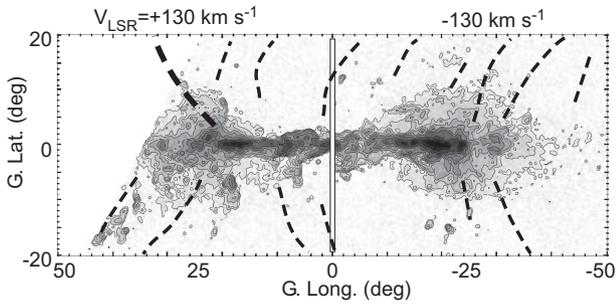}  
\end{center}
\caption{Mosaic HI channel maps at $\vlsr=\pm 130$ \kms produced from the GASS HI survey (McClure-Griffiths et al. 2009; Kalberla et al. 2010). Note the absence of the HI halo inside $|l|<\sim 20\deg$, making an HI hole, and HI spurs making the wall of the hole, approximately coinciding with the radio spurs indicated by dashed lines. }
 \label{chpm130km}  
\end{figure}

Figure \ref{chpm130km} shows mosaic channel maps at $\pm 130$ \kms produced from the HI Galactic All-Sky Survey (GASS; McClure-Griffiths et al. 2009; Kalberla et al. 2010). Symmetric HI spurs emanate from the tangential directions of the 3-kpc ring at $l\sim \pm 20\deg$. Using the GASS HI survey data, we measured the brightness of the wall to be $T\sim 1-2$ K, less bright than the main disk, but the latitudinal extension is as large as $b \sim \pm 10\deg$, reaching altitudes $Z\sim \pm 1.3$ kpc.

In the figure we indicate the radio continuum spurs by the dashed lines, which show the North Polar Spur (NPS), NPS-West, South Polar Spur (SPS), and SPS-West as identified by Sofue (2000). The spurs are commonly traced in galactic surveys of the synchrotron radio emission (Haslam et al. 1982; Reich et al. 2001; Planck Collaboration et al. 2016).  

\begin{figure} 
\begin{center}    
\vskip -5mm \includegraphics[width=7cm]{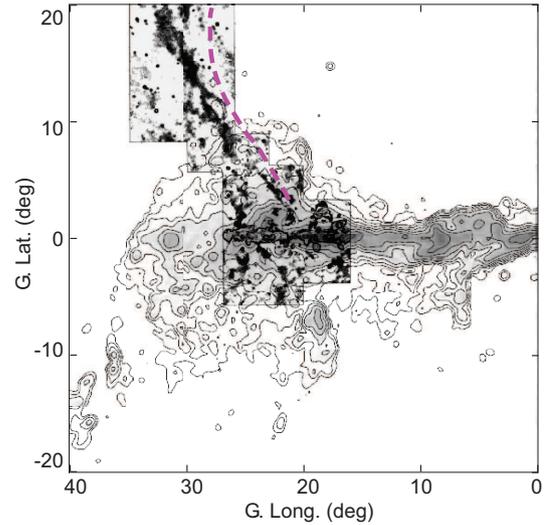}   
\end{center}
\caption{Overlay of the 1.4 GHz continuum spur (Sofue and Reich 1979) on the HI channel map at $\vlsr=+130$ \kms. The dashed line shows the X-ray R7 band (1.4 keV) ridge after correction for the extinction (Sofue 2015) as made from the ROSAT data (Snowden et al. 1997).}
 \label{overlay}  
\end{figure}

Figure \ref{overlay} enlarges the same at $\vlsr=130$ \kms, where we overlay the background-filtered 1420 MHz radio continuum map taken from Sofue and Reich (1979). Also shown is the X-ray ridge in the R7 (1.4 keV) band as obtained from a ROSAT image (Snowden et al. 1997), which coincides with the radio NPS, located slightly inside. The figure demonstrates a remarkable coincidence of the NPS with the 3-kpc HI ring and the wall of the HI hole.

  \section{Vertical Density Profile}
  
\begin{figure} 
\begin{center}   
\includegraphics[width=7cm]{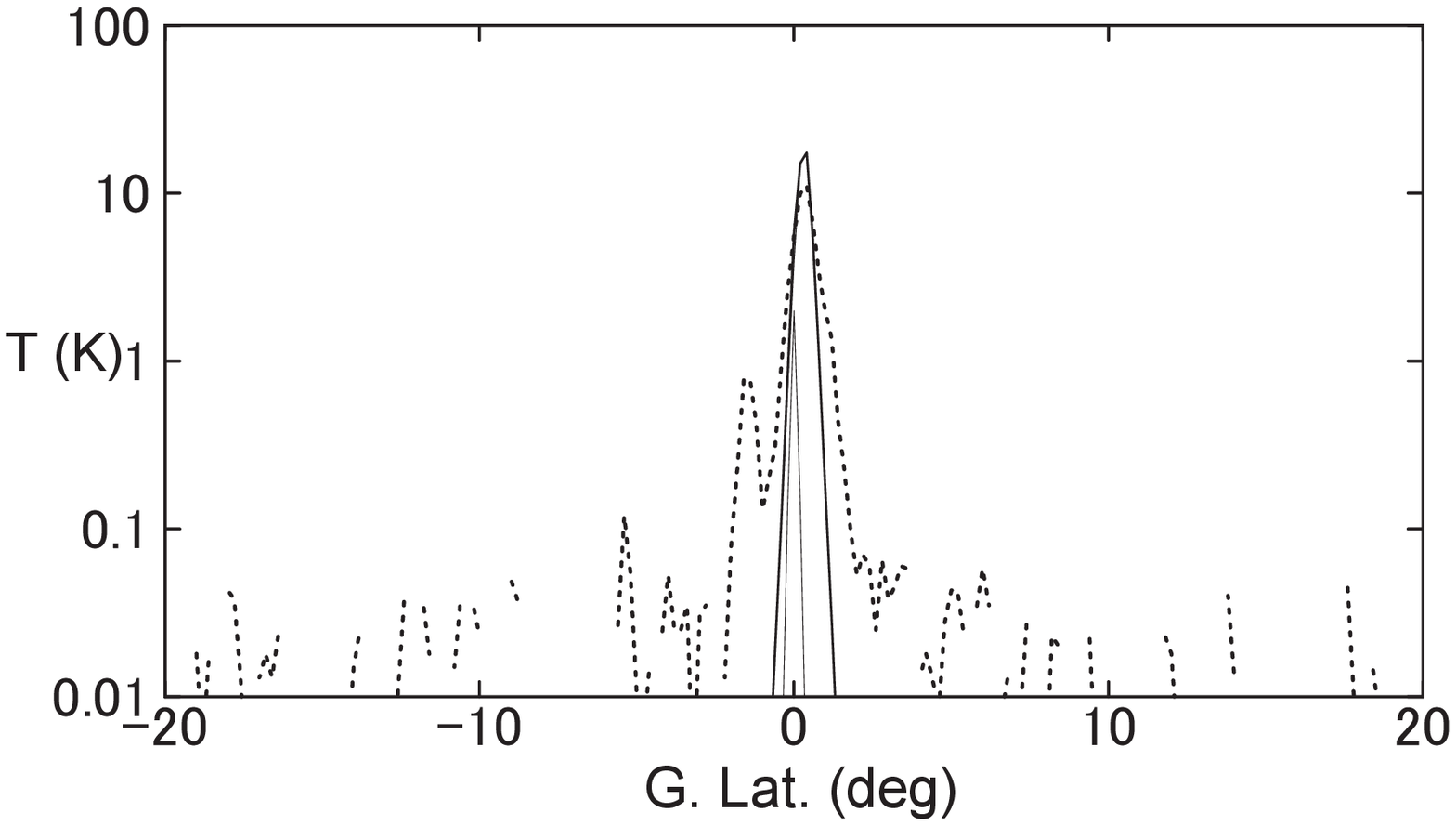}  \\ 
\vskip -38mm $V_{\rm term}=150$ \kms \\
\vskip 33mm\includegraphics[width=7cm]{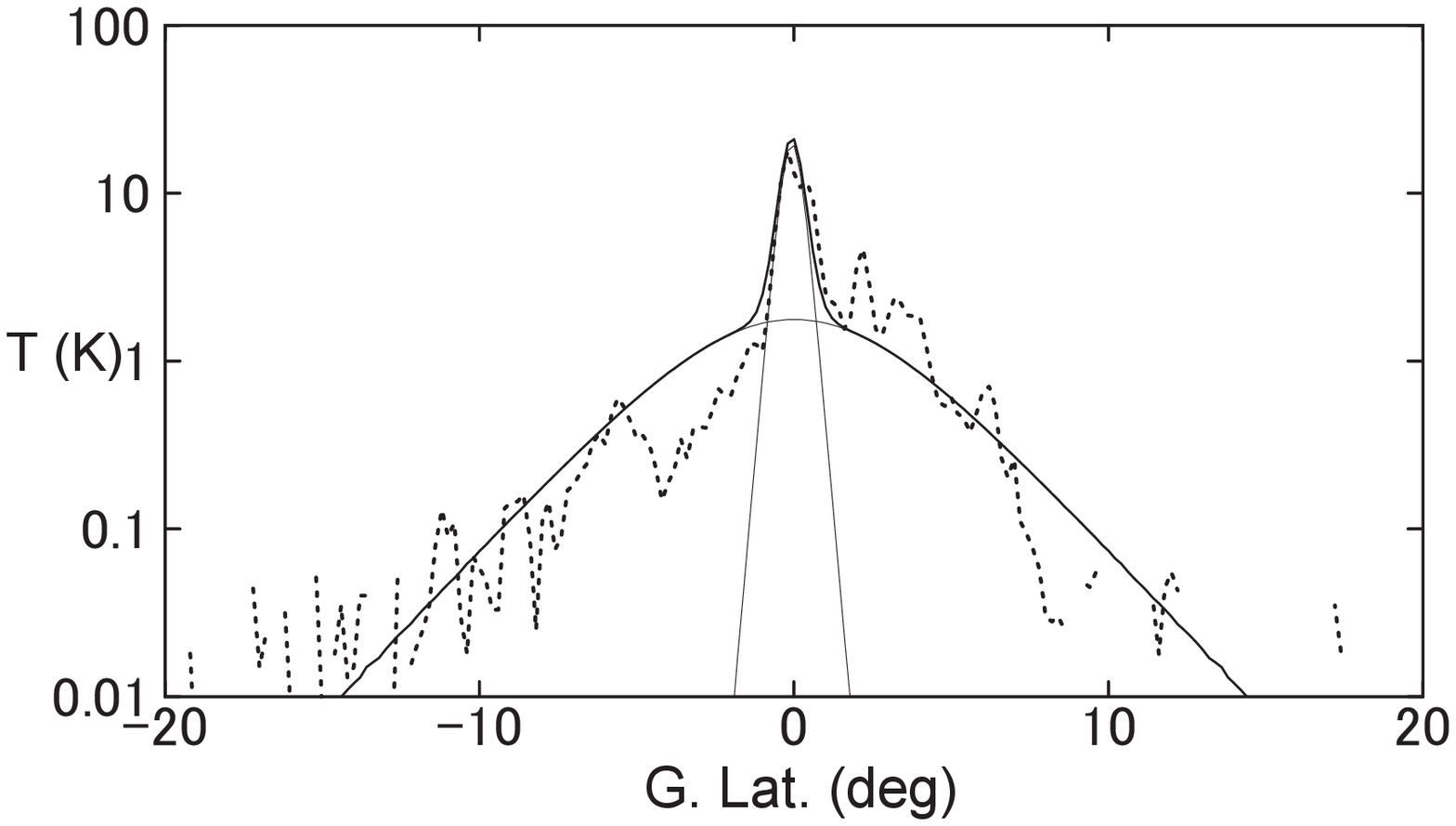}  \\ 
\vskip -38mm $V_{\rm term}=130$ \kms \\ 
\vskip 33mm
\end{center}
\caption{Latitudinal HI intensity ($T_{\rm B}$) profile at terminal velocities $V_{\rm term}=150$ and 130 \kms.
The vertically extended feature is absent at $< 130$ \kms, showing the HI hole inside $l\sim 20\deg$. 
}
 \label{bprof}  
\end{figure}

 Figure \ref{bprof} shows vertical ($b$-directional) cross sections of the HI brightness $T$ at different radial velocities, which show two components with narrow and wide scale heights. 
 We fit the distribution by two gas layers in gravitational equilibrium in the $Z$ direction expressed as 
 \begin{equation} 
\rho =\rho_0 \ {\rm sech}^2 (Z/h)
 \label{sech}
 \end{equation}
 where $h$ is the hydrostatic scale height of the layer (Spitzer 1942). Accordingly, the brightness temperature of the HI emission at a given $\vlsr$ is expressed by
 \begin{equation}
 T =\Sigma_i\ T_i\ {\rm sech}^2 ( Z/{\zhalf} _i), 
 \label{sechT}
 \end{equation}
 where $T_i$ and ${\zhalf} _i$ are the temperature at the midplane and the scale height of the $i$-th components, respectively. Here, $i=1$ and 2 represent the disk and halo components, respectively.

By the $\chi^2$ method we determined $T_i$ and ${\zhalf}_i$ at each longitude. In the inner region with terminal velocity higher than $\vlsr>150$ \kms, the $b$ profiles are fitted by one-disk with small scale height. On the other hand at lower velocities than $\sim 130$ \kms, or at larger longitude than $l \sim 20\deg$, the $b$ profiles are fitted by two components with narrow and wide scale heights. The midplane temperature and scale heights toward the 3-kpc ring are determined to be $T_1=22$ K and ${\zhalf}_1=0\deg.56$ (74 pc) for the disk, and $T_2=5.4$ K and ${\zhalf}_2=3\deg.2$ (450 pc) for the halo component. The halo component is thus well fitted by the sech$^2$ function, indicating that the gas is in gravitational equilibrium. The tight disk component, having ${\zhalf}_1 \sim 74$ pc, represents the main HI disk in the current galactic studies (e.g. Nakanishi and Sofue 2003, 2005, 2015), corresponding to a velocity dispersion of $\sigma_1\sim 5-10$ \kms.

The halo component, showing large scale height of ${\zhalf}_2 \sim 450$ pc, has an order of magnitude less intensity compared to the disk, and is well represented by the hydrostatic equilibrium profile. In order for the gas to be vertically extended to this height, the velocity dispersion must be 
 as large as $\sigma_2\sim 12-25$ \kms.
The HI halo can be traced up to $b\sim \pm 10\deg$ above the detection limit of $T\sim 0.01$ K, indicating that the outskirt is extending to height of $Z\sim \pm 1.3$ kpc

\section{Volume Density Map by Tangent-Circle Method (TCM)}
\label{secmethod}

The column density of HI and H$_2$ gases is related to the velocity integrated intensity of the brightness temperature $T_i$ by
\begin{equation}
N_i=X_i\ \int T_i dv,
\end{equation}
and to the local volume density $n_i$ by
\begin{equation}
n_i=X_i T_i {dv \over dr},
\end{equation}
where $X_i$ is the conversion factor for HI ($i=1$) and CO ($i=2$). 
 We here consider the gas distribution along the tangent circle, which traces the gas having radial velocities equal to the terminal velocities.
Using the relation $r=R_0\ {\rm cos}\ l$, $R=r\ {\rm tan}\ l=R_0 \ {\rm sin }\ l$, and $dl/dr= 1/R_0 {\rm sin}\ l$ at the tangent points, we obtain
\begin{equation} 
n_i=X_i T_i \left({V_0 \over R_0} - {dV \over dR} \right)\ {\rm cot}\ l.
\label{eqTCM}
\end{equation} 
We here assume that the HI gas is optically thin, as we are interested in the halo gas.
 
We can thus transform the observed brightness temperature $T_i$ on the LV diagram along the terminal velocity ridge into the local volume density $n_i$ as a function of the radius $R$. We call this method the tangent-circle method (TCM). The method can avoid the degenerate depth problem at the tangent points when calculating the volume density from the intensity (column density).

We applied the TCM to the HI line data from the GASS (McClure-Griffiths et al. 2009; Kalberla et al. 2010), Leiden-Argentine-Bonn (LAB) HI survey (Kalberla et al. 2005), and CO line data from the Columbia galactic plane CO-line survey (Dame et al. 2001). In order to calculate $dV/dR$, we adopt the most recent Galactic constants ($R_0=8$ kpc and $V_0=238$ \kms: Honma et al. 2015) and rotation curve (Sofue 2016). The conversion factors are taken to be $X_{\rm HI}=1.82\times 10^{18}$ H cm$^{-2}$ and $X_{\rm CO}=2.0\times 10^{20}$ H$_2$ cm$^{-2}$ $=4.0\times 10^{20}$ H cm$^{-2}$.

\begin{figure*}  
\begin{center}    
\vskip -2mm \includegraphics[width=12cm]{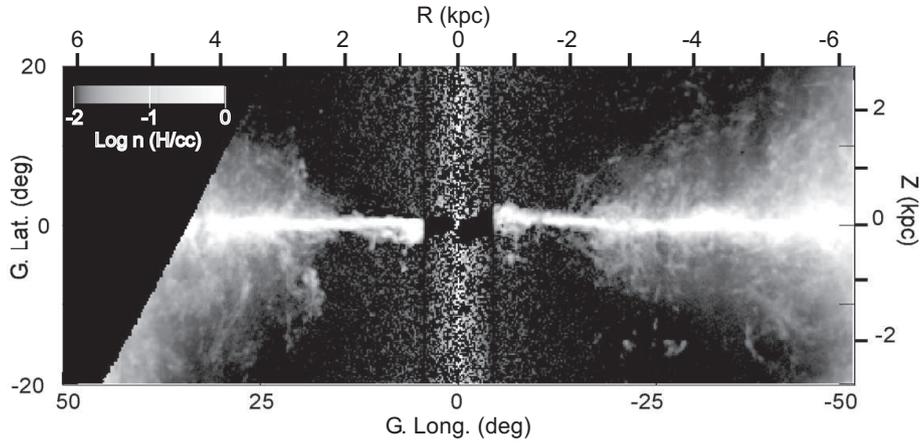}
\end{center} 
\vskip -2mm
\caption{
 HI volume density map in the $(R,Z)$ plane using the tangent-circle method (TCM). The map outside $\pm \sim 5$ kpc is not accurate (see the text). 
} 
 \label{MapLog}   
\end{figure*}

 Figure \ref{MapLog} shows the distribution map of the H density, $n_{\rm H}$, along the tangent circle in the $(l,b)$ and $(R,Z)$ plane. Thereby, the TCM was applied to the terminal velocity ridge approximated by a straight line in the LV diagram parallel to the tangent ridge around $l\sim 20\deg$. By this approximation, the map is not accurate at $|l|>\sim 40\deg$ and $|l|<\sim 5\deg$. The map is essentially the same as the column density map obtained by Lockman and \MG (2016). However, it gives the volume density distribution, which exhibits a sharper and clearer-cut wall thanks to the pin-pointing nature of the solar-circle gas by the TCM.

\begin{figure} 
\vskip -5mm
\begin{center}   
(a) Radio conti. excess at $b=+5\deg$\\   
\includegraphics[width=8cm]{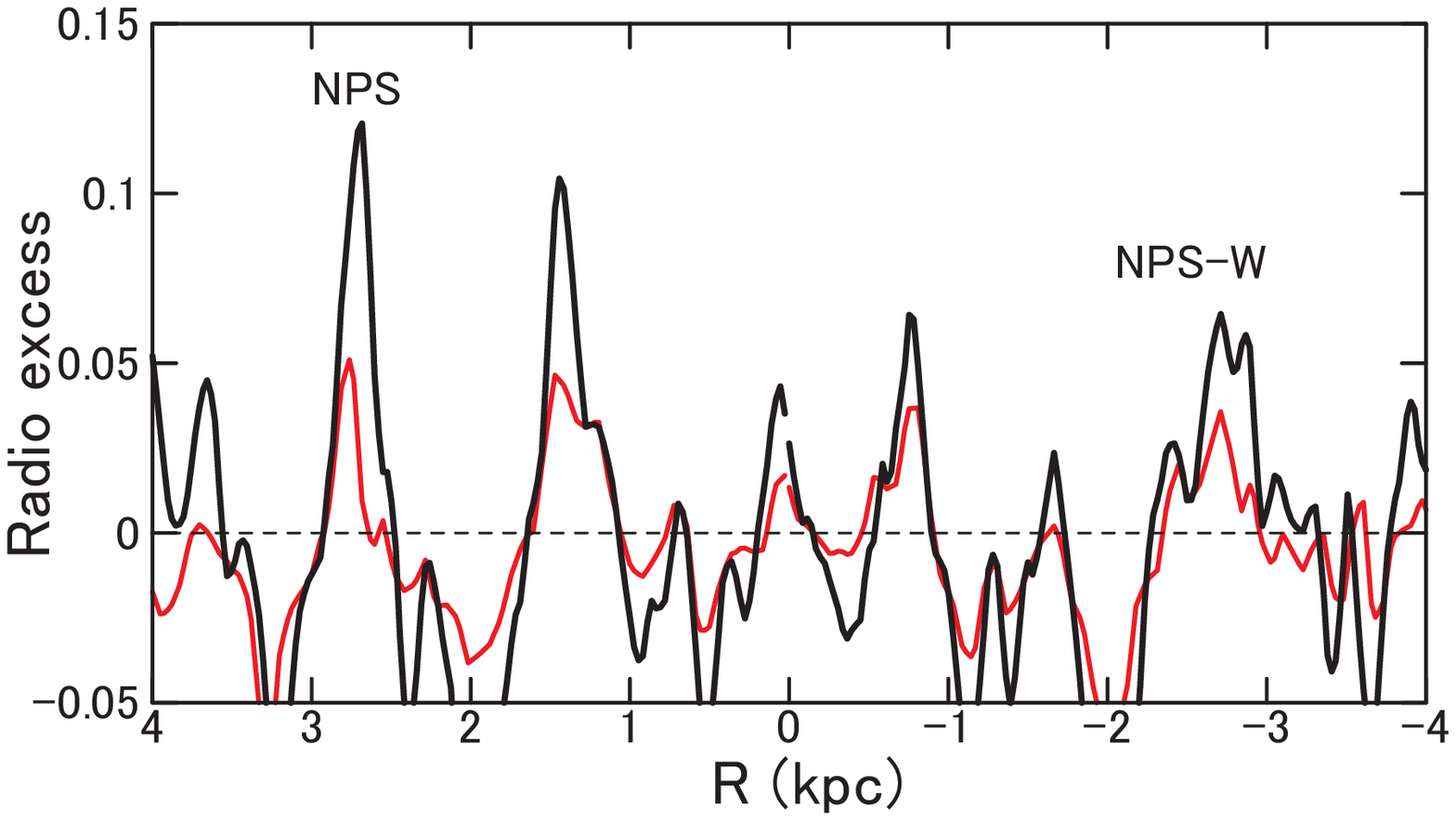} \\ 
\vskip -1.5mm
(b) HI at $b=+3\deg$\\ 
\includegraphics[width=8cm]{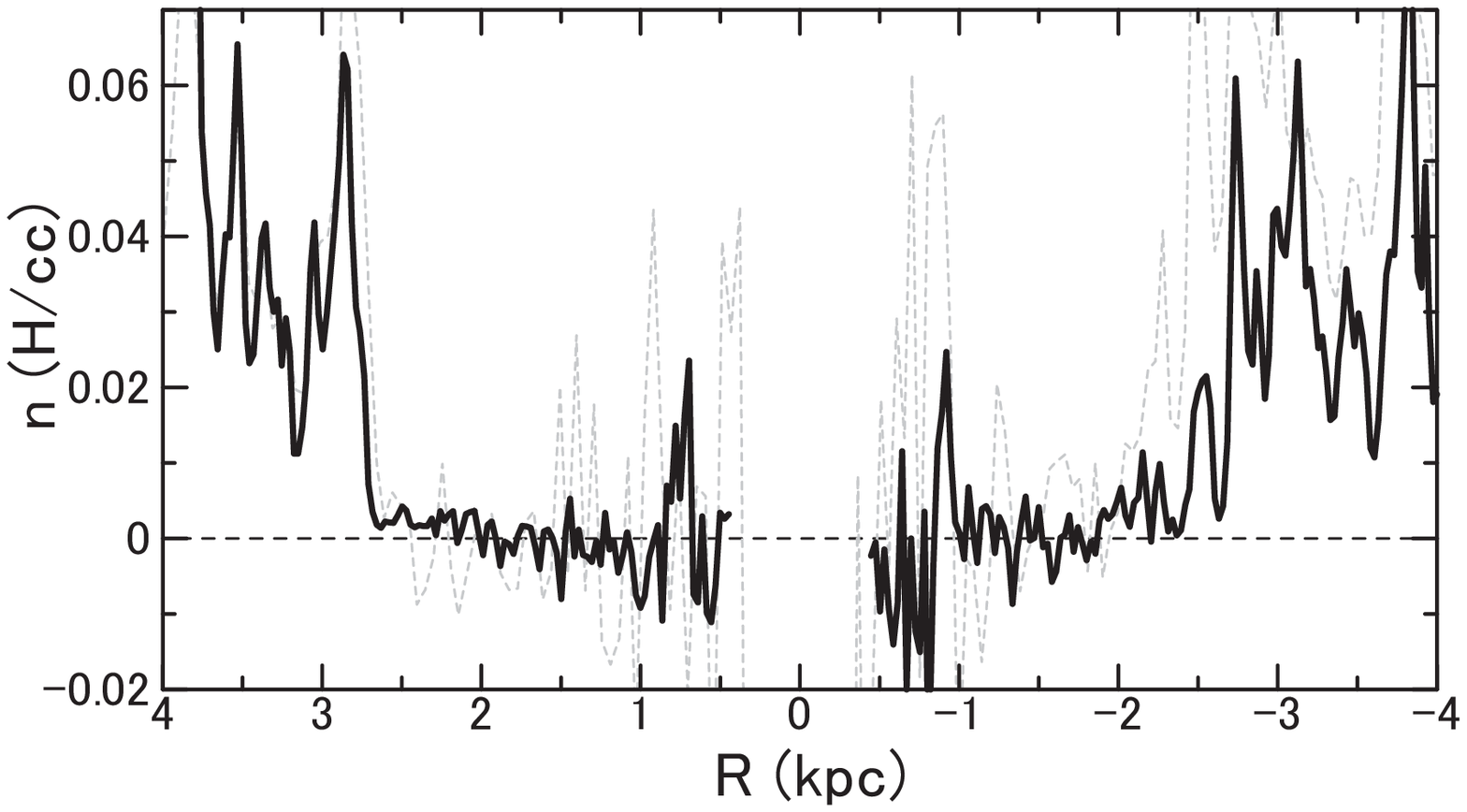} \\
\vskip -1.5mm
(c) HI and CO at $b=0\deg$\\  
\includegraphics[width=8cm]{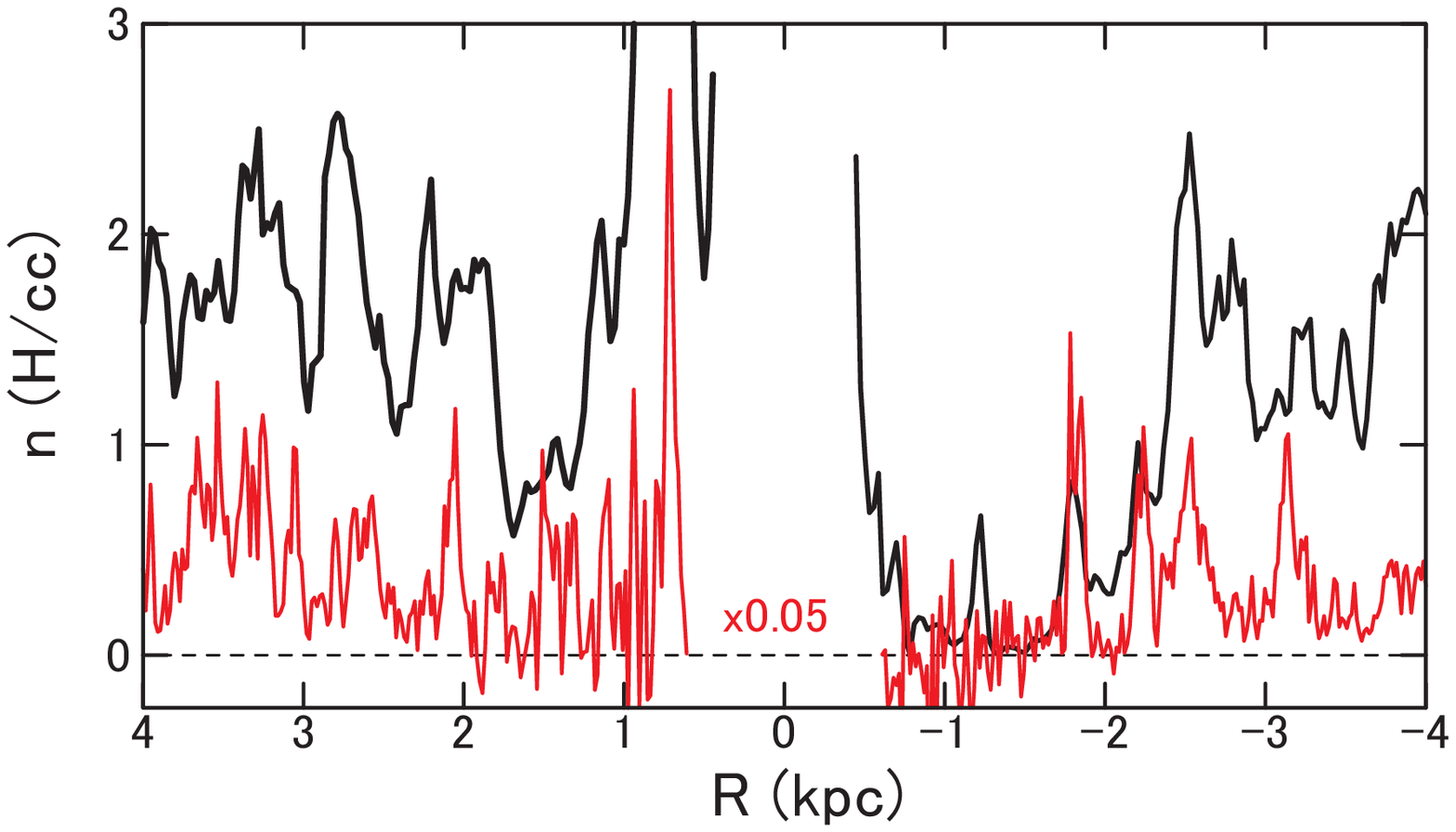}
\end{center}
\vskip -1.5mm
\caption{
(a) Radio continuum excess at $b=+5\deg \ (\sim 700$ pc). Black = 408 MHz; red = 1420 MHz.  Negative $R$ stands for negative longitude. 
(b) Volume density $n$ of H atoms plotted against $R$ at $b=3\deg \ (\sim 400$ pc). Full = GASS; grey dash = LAB. 
(c) Same, but at $b=0\deg$ for HI (black) and CO (red, $n \times 0.05$).  
} 
 \label{rprof}   
\end{figure}

In figure \ref{rprof} we compare the HI density profiles at fixed latitudes with those of the CO-line and radio continuum emissions. Panels (a)-(c) show:\\
(a) Radio continuum excess at 408 (Haslam et al. 1982) and 1420 MHz (Reich et al. 2001) at $b=5\deg$. The radio excess is defined by $ \conex=\Tb/\Tav-1,$, where $\Tb$ is the brightness temperature and $\Tav$ is the smoothed intensity in an area $\delta l \times \delta b = 5\deg \times 1\deg$ around each data point.\\
 (b) HI density profiles at $b=+3\deg$, where the full and dashed lines show the results for the GASS and LAB data, respectively. The difference between the two results are due to the difference in the beam widths as well as the noise levels. The increasing scatter toward the GC is due to the cot $l$ effect in equation \ref{eqTCM} on the errors. \\
 (c) HI and molecular hydrogen densities along the galactic plane at $b=0\deg$ from GASS and Columbia CO survey data. 

The HI crater is now evident in the profiles at $b=+3\deg \ (\sim 400$ pc). It is remarkable that the crater edges exactly coincide with the radio continuum peaks of the NPS and NPS-W. About the same HI profiles were obtained at $b\sim 3 - 10\deg$ as well as at negative latitudes except for the decreasing density with $|b|$. 

The HI density inside the hole is as low as $n_{\rm HI}\le \sim 5\times 10^{-3}\ \Hcc$. At the edge of the hole, the intensity suddenly increases to a sharp peak at $R=2.9$ kpc, making a clear-cut wall. The peak density in the wall is $n_{\rm HI}\sim 0.07\ \Hcc$. The full radial width at half-density of the wall is measured to be $\Delta R \simeq 0.15$ kpc in the first quadrant ($l\ge 0\deg$). 

The HI profile in the galactic plane ($b=0\deg$) is similar, but much milder. The intensity peaks representing the 3-kpc ring appear at $R\sim 2.8$ kpc, coinciding with the HI walls. The peak HI density is $n_{\rm HI} \sim 2.5 \  \Hcc$, and the full width of the ring is $\Delta R \simeq 0.3$ kpc. It is found that the peak of the 3-kpc ring is more evident at negative longitudes.

The molecular disk is clumpy (figure \ref{rprof}(a) red lines), while it still exhibits peaks at $R\sim 2.7$ kpc with density of $n_{\rm H_2}\sim 13\ \Hcc$. The molecular fraction (molecular density/total density) in the peaks (wall) is $f_{\rm mol}\simeq 0.8$, lower than that in the surrounding regions with $f_{\rm mol}\simeq 0.9$, in agreement with the variation of molecular fraction in the inner Galaxy obtained by Sofue and Nakanishi (2016).  

Using the measured parameters we calculated the total mass of the 3-kpc ring, assuming a perfect circle around the GC with a constant peak density. We also calculated the kinetic energy, assuming that the expansion velocity is $V_{\rm expa}=53$ \kms. The derived quantities are listed in table \ref{tabpara}.

\begin{table*}  
\caption{Parameters of the 3-kpc HI crater.}  
\begin{center}
\begin{tabular}{lllllll} 
\hline 
& CO ring & HI ring & HI wall & HI hole & Total \\
 \hline 
 Radius$^\dagger$, $R$ (kpc) & 2.7  & 2.8  & 2.9  & 2.8&---   \\
 Scale height, $h$ (kpc)& 0.07 & 0.07 & 0.4 & $>1$ &---\\
 Width, $\Delta R$ (kpc) & 0.15  & 0.3 & 0.15 & ---  &---\\
 Density, $n$ ($\Hcc$) & 13 & 2.5 & 0.1 & $<5\times 10^{-3}$ &---\\
 Mass, $M$ ($\Msun$) & $1.1\times 10^8$ & $4.3\times 10^7$ &  $0.6\times 10^7$ & ---&  $1.6\times 10^8$\\
 Expa. vel$^{\ddagger}$, $V_{\rm expa}$ (\kms) & 53 & 53 & 53 & ---&--- \\
 Kin. Energy, $E_{\rm kin}$ (erg) & $3.1\times 10^{54}$ & $1.3\times 10^{54}$ & $1.5\times 10^{53}$ &---& $4.7 \times 10^{54}$   \\
\hline     
$\dagger$ for $R_0=8.0$ kpc; $\ddagger$  Cohen and Davies (1976)
\end{tabular}  
\end{center}  
\label{tabpara}
\end{table*}   

\section{A Model for the HI Crater}   

We now examine if the crater structure in the HI halo can be explained by sweeping of the halo gas by an explosive event in the GC based on the bipolar-hyper-shell (BHS) model of the NPS (Sofue 2000; Sofue et al. 2016). The propagation of a shock wave from the Center is calculated using the Sakashita's (1971) method to trace radial ray paths of an adiabatic shock wave.

The unperturbed gas disk and halo are assumed to be composed of stratified layers with density distributions represented by the hydrostatic equilibrium in the $Z$ direction (Spitzer 1942). It is assumed that the disks are further embedded in an intergalactic gas with uniform, low-density gas. We express the density distribution as
\begin{equation}
\rho=\Sigma \rho_i \ {\rm sech}^2 ( Z/{\zhalf}_i).
\label{eqrho}
\end{equation}
Here, $i=1,\ 2,$ and 3 represent the disk, HI halo, and a constant background. We here take $\rho_1=1,\ \rho_2=0.1,$ and $\rho_3=10^{-5}$ H cm$^{-3}$, and  ${\zhalf}_1=50,\ {\zhalf}_2=500$ pc, and ${\zhalf}_3=\infty$. 

\begin{figure} 
\begin{center}    
\vskip -5mm
\includegraphics[width=6.5cm]{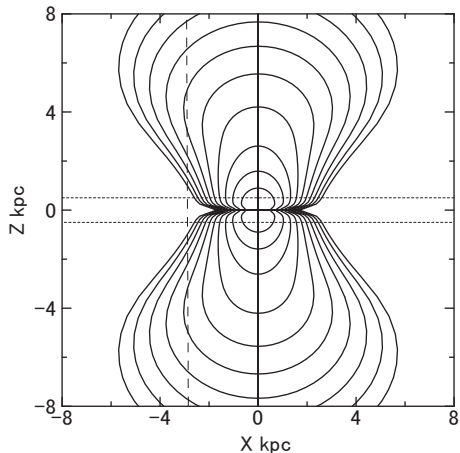}      
\end{center}
\vskip -5mm
\caption{Propagation of the shock front in the disk and halo. The fronts are drawn every 1 My.
 }
 \label{model}  
\end{figure}

Figure \ref{model} shows the calculated result for an initial injection energy $E_0=1.8\times 10^{55}$ erg. The shock front is drawn in the $(R,Z)$ plane every 1 My up to 10 My. As the shock wave expands, the front shape becomes elongated in the vertical direction due to the steep pressure gradient toward the halo. As the shock wave is blown off into the halo, the front shape gets dumbbell shaped, making a BHS. The dumbbell's equator is sharply pinched by the dense disk at the galactic plane. At elapsed time of $t\sim 10$ My, the BHS front approximately mimics the NPS, SPS, NPS-W and SPS-W. 

The expanding velocity of the front at intermediate latitudes $b \sim 10-20\deg$ corresponding to the main NPS ridge is $\ve\sim 300$ \kms. This velocity is coincident with the required velocity to heat the shocked gas to a temperature $\sim 10^7$ K responsible for the observed X-ray emission in the NPS (Snowden et at al. 1997; Sofue et al. 2016). Note, however, the shock-heated gas inside the shock front is no more neutral (HI), but is ionized to X-ray temperatures, and is not observed in the HI line emission. The HI wall outside the shock front is in a pre-shock compression stage, and is observed as the expanding HI ring. The ring's expansion is still slow, as observed to be expanding at $\ve \sim 50$ \kms, and approximately obeys the normal galactic rotation.

\begin{figure} 
\begin{center}   
\vskip -5mm
\includegraphics[width=8cm]{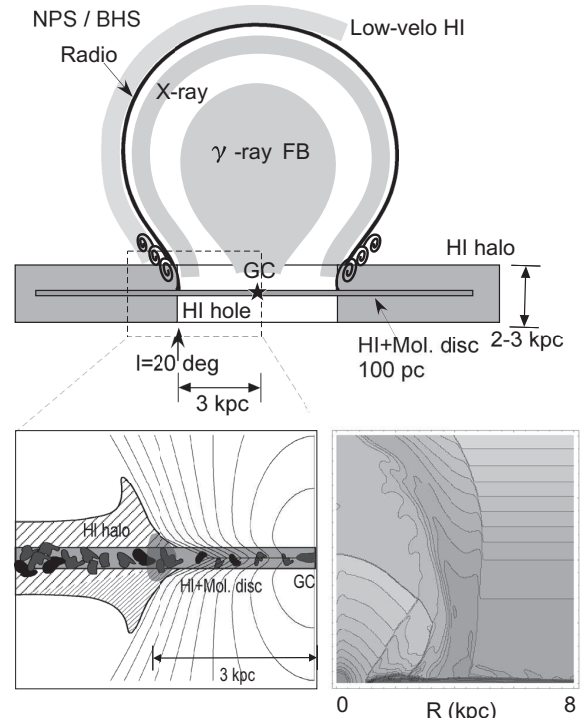}
\end{center}
\vskip -2mm
\caption{ 
Schematic view of the galactic crater. The bottom-right panel is the hydrodynamic BHS model at 10 My showing the density in logarithmic scale reproduced from Sofue et al. (2016). 
}
 \label{sketch}  
\end{figure}

In figure \ref{sketch} we schematically summarize the view about the radio and X-ray NPS (BHS), 3-kpc ring, HI hole and wall, HI halo, and the dense main (HI+H$_2$) disk. The enlarged illustration is drawn by referring to the hydrodynamical BHS model, showing that the dense disk is kept unperturbed inside the global front. The view is consistent with the hydrodynamic simulation shown by the inserted reproduction from Sofue et al. (2016) at $t=$ 10 My. In the present model, the Fermi Bubble is considered to be a younger object of a few My related to the innermost expanding ring in the GC (Sofue 2017). 

\section{Discussion}

\subsection{Summary} 

We derived the volume density map of HI gas along the tangent circle, and showed that the HI halo has a large crater-shaped structure around the GC. Thereby, we developed the TCM for pin-pointing the solar-circle gas without suffering from the kinematically degenerate depth problem in the tangent-circle direction.

The wall of the HI crater positionally coincides with the brightest emission ridge of the NPS and the tangential direction of the 3-kpc expanding ring.  While the HI hole inside the crater is almost empty, the outside halo is not disturbed, being kept in hydrostatic equilibrium (sech$^2 Z/h_2$) with a scale height $h_2\sim 0.45$ kpc. 

The origin of the crater structure is explained by a shock wave model based on the giant explosion hypothesis at the GC. Figure \ref{sketch} summarizes the observed structures in HI, CO, radio, X-ray, and $\gamma$ rays.

\subsection{Some difficulties in the models} 

In our model the shock front expands into the halo, whereas the dense galactic disk is not strongly disturbed. In order to accelerate the 3 kpc ring to the observed velocity, Sanders and Prendergast (1974) had to assume an energy as large as $\sim 3\times 10^{58}$ erg, but such a huge explosion totally destroyed the disk. Therefore, the hydrodynamical models cannot reproduce the expanding motion of the 3-kpc ring. Such a local acceleration could be possible by refraction of waves transmitting the halo and focusing onto the disk as suggested by the MHD wave propagation model (Sofue 1977, 1984).

This kinematics problem is not encountered by the bar hypothesis (e.g. Binney et al. 1991), where no energetic explosion is required to produce the non-circular motion. On the other hand, the largely extended vertical structure of the HI wall and the empty hole in the HI halo might not be easy to explain by the bar. 

\subsection{Low-velocity HI shell}

It has been suggested that low-velocity HI gas at $|\vlsr| <\sim 50$ \kms apparently surrounds the NPS (Heiles et al. 1980). However, it is not clear if the HI is indeed related, because the low velocity HI map is full of bright, often much brighter, local HI shells and filaments, making it difficult to confirm their true relation.

If the HI outer shell is indeed associated with the NPS, the following scenario would be possible. The swept-up gas by the BHS is compressed to form high-temperature X-ray gas, while the densest front cools down to neutral gas and drops toward the disk. Since the snow-plowed gas has not enough angular momentum, the galactic rotation is somehow canceled by the dropping gas. A dynamical model taking account of the cooling of compressed gas to HI temperature remains as a subject for future consideration.  

\section*{Acknowledgements}
 The author is grateful to the authors of the GASS and LAB HI surveys and the Columbia CO survey for providing us with the archival data cubes.  

\def\r{ \bibitem[]{} }

\end{document}